\documentclass{IEEEtran}
\usepackage{cite}
\usepackage{amsmath,amssymb,amsfonts}
\usepackage{algorithmic}
\usepackage{graphicx}
\usepackage{epstopdf}
\usepackage{textcomp}
\usepackage{subcaption}
\usepackage{array}

\def\BibTeX{{\rm B\kern-.05em{\sc i\kern-.025em b}\kern-.08em
    T\kern-.1667em\lower.7ex\hbox{E}\kern-.125emX}}

\begin{document}

\IEEEpubid{\begin{minipage}[t]{\textwidth}\ \\ \\ [2pt]
        \footnotesize{This work has been submitted to the IEEE Access for possible publication. Copyright may be transferred without notice, after which this version may no longer be accessible.}
\end{minipage}} 

\title{Decoupling and Matching Strategies for Compact Antenna Arrays}
\author{F. Estev\~ao. S. Pereira, Josef A. Nossek,  \IEEEmembership{Life Fellow, IEEE}, and F. Rodrigo P. Cavalcanti
\thanks{``This work was partially funded by Research Support Foundation of Cear\'a State (FUNCAP) and the Higher Education Personnel Improvement Coordination (Capes), and was also supported in part by the Wireless Telecommunications Research Group (GTEL), Department of Teleinformatics Engineering, Federal University of Cear\'a, Brazil.'' }
\thanks{F. Estev\~ao. S. Pereira is with the Wireless Telecommunications Research Group (GTEL), Departament of Teleinformatics Engineering, Federal University of Cear\'a, Brazil (e-mail: estevaosimao@gtel.ufc.br).}
\thanks{Josef A. Nossek is with the Department of Electrical and Computer Engineering, Technical University of Munich, Germany  (e-mail: josef.a.nossek@gtel.ufc.br).}
\thanks{F. Rodrigo P. Cavalcanti is with the Wireless Telecommunications Research Group (GTEL), Departament of Teleinformatics Engineering, Federal University of Cear\'a, Brazil (e-mail: rodrigo@gtel.ufc.br).}
}

\maketitle

\begin{abstract}
Antenna arrays have been used in various applications and have become an important tool to achieve high spectral efficiency in wireless communications. Its use brings to the communications system an increase in performance in terms of capacity and reliability. Recently one of the main communications architectures that makes use of antenna arrays is the Multiple-Input Multiple-Output (MIMO) technology. MIMO technology has been applied in antenna arrays composed of elements in the order of tens to hundreds. In this way, it is necessary to use the structure of compact antennas that offer all the necessary robustness to the applied project. In this context, it is important to revisit the concepts of mutual coupling and impedance matching among antenna elements in an array. This paper proposes and evaluates three strategies of joint decoupling and impedance matching networks(DMN) for antenna arrays. The first method called DMN with Lumped Elements (DMN-LE) performs the decoupling and impedance matching steps with capacitors and inductors. The second method is called the Ring Hybrid (DMN-RH). It is utilizes a microstrip line in the ring form. With this approach is achieved first the decoupling followed by impedance matching steps. The third method is called Networkless Decoupling and Matching (NDM). It brings a concept of decoupling without the presence of a network itself. A comparison of the methods is performed both analytically and via computer simulations. We conclude that the third method, networkless one, is an interesting new alternative approach.
\end{abstract}


%\footnote{This work has been submitted to the IEEE for possible publication. Copyright may be transferred without notice, after which this version may no longer be accessible.}

\section{Introduction}
\label{sec:introduction}
\IEEEPARstart{A}{ntenna} arrays have been used in various applications and have become an important tool to achieve high spectral efficiency in wireless communications. Its use brings to the communications system an increase in performance in terms of capacity and reliability.

Recently one of the main communications architecture that makes use of antenna arrays is the Multiple-Input Multiple-Output (MIMO) technology. It has become an important technique in wireless communications systems. MIMO employs transmit and receive antenna arrays so as to enable performance gains from both spatial diversity and spatial multiplexing \cite{Cit1}, \cite{Cit2}, \cite{Cit3}, \cite{Cit4}. In this context, it is important to revisit the concepts of mutual coupling and impedance matching among antenna elements in an array. 

The mutual coupling is the effect of interaction among the elements of antenna arrays which can generate losses in gain and efficiency of the antenna system if not taken properly into account. According to \cite{Cit5} this effect basically depends on the radiation characteristics of each antenna, the relative separation among antennas and also the relative orientation of each antenna. 

Impedance matching aims at maximum power transfer between amplifiers of the communication system and the antenna array. The correct operation of an antenna system in a given frequency range does not depend exclusively on its own impedance characterization but on its combination with the impedance properties of the communications system itself. 

Nevertheless, performing impedance matching in an antenna array can often make the antenna system structure, as a whole, more complex, requiring larger dimensions for its implementation. In this way, efficient and, at the same time, compact arrays and impedance matching techniques are desired.

This paper proposes and evaluates three strategies of joint decoupling and impedance matching networks (DMNs) in an array of 2 dipole antennas. From the multiport communication theory presented in \cite{Cit3} the appropriate equations for the proposed DMN are derived. We offer an analysis of the proposed methods in order to identify a compact and low complexity solution for the realization of the decoupling and impedance matching steps in a single structure. 
  
The first method called DMN with Lumped Elements (DMN-LE) performs the decoupling and impedance matching steps by a concentrated circuit with capacitors and inductors.
The second method is called the Ring Hybrid (DMN-RH). It is modeled with a microstrip line in a ring format. With this modeling the ring line segments are designed so that the Ring Hybrid ports for the amplifiers are decoupled.
The third method is called Networkless Decoupling and Matching (NDM). It brings a concept of decoupling and impedance matching without the presence of a multiport network itself.

A case study will be conducted in order to compare the performance results and obtain a DMN structure capable of providing high efficiency to the antenna array and the communications system.

The rest of paper is organized as follows: Section II brings the general definition and concept of the Decoupling and Matching Network (DMN). The Section III presents the DMN with lumped elements along with an analysis of the system. Section IV presents the DMN with the use of Ring Hybrid, explaining the process of its construction. Section V demonstrates the Networkless decoupling technique along with the impedance matching step. The section VI shows computer simulation results and discusses them. And finally section VII brings the conclusions achieved.     

% ----------------------------------------------------------
% Seção: DMN
% ----------------------------------------------------------
\section{Decoupling and Matching Network (DMN)}

\begin{figure}[h]
    \centering
    \centerline{\includegraphics[width=0.5\textwidth]{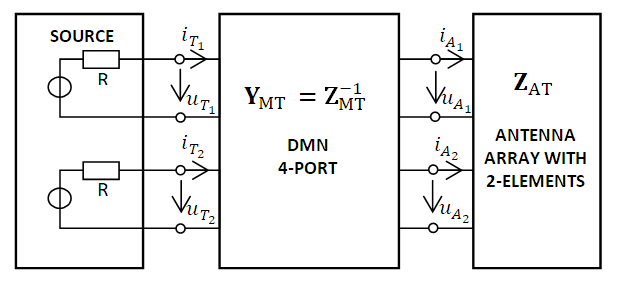}}
    \caption{4-Port DMN block representation. }
    \label{fig:fig1}
\end{figure}

The Decoupling and Matching Network (DMN) is a device capable of performing the decoupling operations between the elements of antenna array and the impedance matching on the output ports. The DMN ports are designed to provide an impedance of $50 \Omega$ on the source side and on the other side the ports provide an equivalent impedance value to the antenna array. The Fig.\ref{fig:fig1} shows the block representation of the 4-port DMN providing decoupled ports matched to the sources each with source impedance R $= 50 \Omega$. 

This diagram represents the transmission configuration of the signal, which will be propagated by the antenna array. In this representation the source comprised of ports $T_{1}$ and $T_{2}$ represents the insertion of the signal that will be transmitted in each port. On the other side of the DMN block, on the ports $A_{1}$ and $A_{2}$, the antenna array block represents its impedance matrix ($\textbf{Z}_{AT}$). 

Equations \eqref{eq:001} to \eqref{eq:004} shows the definition of the voltage and current vectors in the T and A ports.

\begin{gather}
\label{eq:001}
  \textbf{\textit{u}}_{T}
  =
 \begin{bmatrix}
    \textit{u}_{T_1} \\
    \textit{u}_{T_2}  
\end{bmatrix}
\end{gather}

\begin{gather}
\label{eq:002}
  \textbf{\textit{u}}_{A}
  =
 \begin{bmatrix}
    \textit{u}_{A_1} \\
    \textit{u}_{A_2}  
\end{bmatrix}
\end{gather}

\begin{gather}
\label{eq:003}
  \textbf{\textit{i}}_{T}
  =
 \begin{bmatrix}
    \textit{i}_{T_1} \\
    \textit{i}_{T_2}  
\end{bmatrix}
\end{gather}

\begin{gather}
\label{eq:004}
  \textbf{\textit{i}}_{A}
  =
 \begin{bmatrix}
    \textit{i}_{A_1} \\
    \textit{i}_{A_2}  
\end{bmatrix}
\end{gather}

The purpose of this analysis is to obtain the matrix $\textbf{\textit{Z}}_{MT}$ such that the sources achieve power matching and decoupling to the DMN according to \eqref{eq:0041}. This can be achieved with the following derivations in the \eqref{eq:005} to \eqref{eq:009}, 

\begin{gather}
\label{eq:0041}
\textbf{\textit{u}}_{T} =
  R.\textbf{\textit{1}}.\textbf{\textit{i}}_{T}   , \end{gather}
 
\begin{gather}
\label{eq:005}
 \begin{bmatrix}
    \textbf{\textit{u}}_{T} \\
    \textbf{\textit{u}}_{A}  
\end{bmatrix}
  =
  \textbf{\textit{Z}}_{MT}  
 \begin{bmatrix}
    \textbf{\textit{i}}_{T} \\
    -\textbf{\textit{i}}_{A}  
\end{bmatrix}   ,
\end{gather}

\begin{gather}
\label{eq:006}
 \textbf{\textit{u}}_{A} =
  \textbf{\textit{Z}}_{AT}
  \textbf{\textit{i}}_{A}   ,
 \end{gather}

\begin{gather}
\label{eq:007}
  \textbf{\textit{Z}}_{MT}
  =
 \begin{bmatrix}
    \textbf{\textit{0}} & \textbf{\textit{X}}_{1} \\
    \textbf{\textit{X}}_{1} & \textbf{\textit{X}}_{2} 
\end{bmatrix}
\end{gather}

where, 
\begin{gather}
\label{eq:008}
 \textbf{\textit{X}}_{1} = -j\sqrt{R} Re \big\{ \textbf{\textit{Z}}_{AT} \big\}^{-\frac{1}{2}}  
\end{gather} 

\begin{gather}
\label{eq:009}
 \textbf{\textit{X}}_{2} = -j Im \big\{ \textbf{\textit{Z}}_{AT} \big\}^{}.    
\end{gather}

\begin{figure}[h!]
    \centerline{\includegraphics[width=0.3 \textwidth]{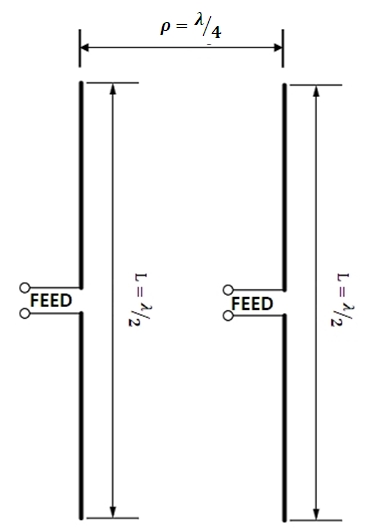}}
    \caption{Dipole Antenna Array with two elements.}
    \label{fig:antenna}
\end{figure}

The antenna array that will be discussed in this article is presented in the Fig.\ref{fig:antenna}. This array consists of two half-wavelength dipole antennas where the distance between them is given by $\rho$. The structure feed is also represented in the figure, located in the center between the $\lambda/4$ segments.

% ----------------------------------------------------------
% Seção: DMN with Lumped Elements
% ----------------------------------------------------------
\section{DMN with Lumped Elements (DMN-LE)}

The DMN-LE performs the decoupling between the elements of the antenna array and the impedance matching between the array and the source through lumped elements. Lumped elements to perform impedance transformation are widely used in microwave systems \cite{Cit6}, \cite{Cit7}, \cite{Cit8}, \cite{Cit9}, \cite{Cit10}. However in certain situations its use can make the antenna structure very complex, requiring several stages in cascades to reach the desired impedance \cite{Cit9}, \cite{Cit10}.

\subsection{Design of DMN Lumped Elements}
\begin{figure}[h!]
    \centerline{\includegraphics[width=0.5\textwidth]{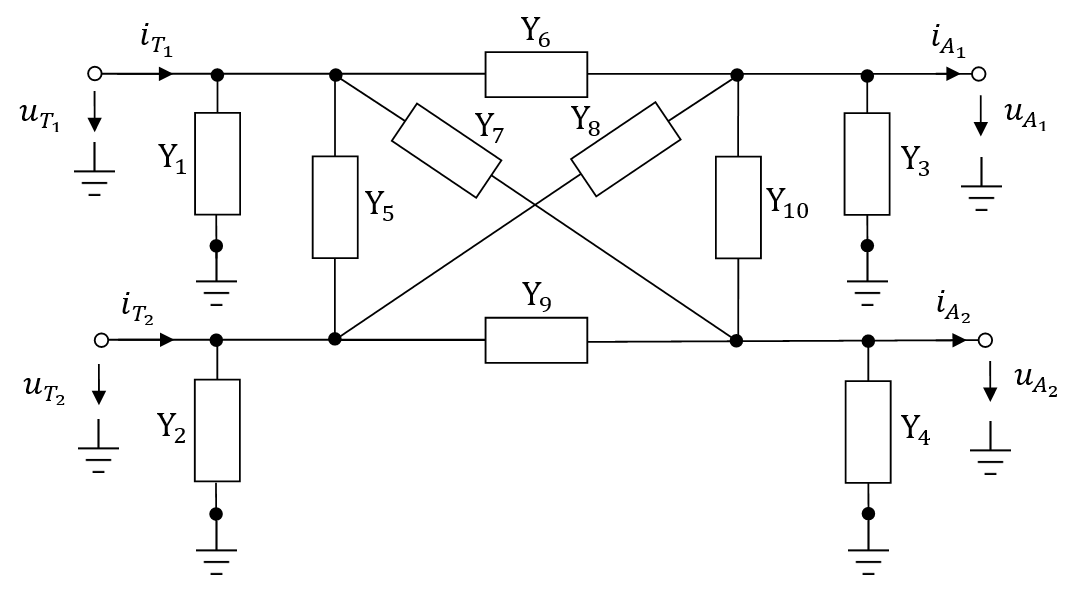}}
    \caption{Representation of the components to DMN-LE block.}
    \label{fig:fig2}
\end{figure}

For the detailed design of DMN-LE it is convenient to work with the admittance matrix $\textbf{\textit{Y}}_{MT}$. Its representation will be referred to as a general 4-port implemented with $Y_1$ to $Y_{10}$ according to Fig.\ref{fig:fig2}. For this diagram the ports $u_{T_1}$ and $u_{T_2}$ represent the input ports of the DMN-LE while $u_{A_1}$ and $u_{A_2}$ represents the output ports connecting the antenna array. The DMN-LE is designed to operate at a center reference frequency $f_{r}$.

The DMN-LE is initially characterized by $\textbf{\textit{Y}}_{MT}$ and in order to perform the modeling of each element of the DMN-LE it is necessary to know the impedance matrix $\textbf{\textit{Z}}_{AT}$ of the antenna array.
For the derivation of matrix $\textbf{\textit{Y}}_{MT}$ \eqref{eq:1}, \eqref{eq:11} and \eqref{eq:12} are used. 

\begin{gather}
\label{eq:1}
  \textbf{\textit{Y}}_{MT}
  =
 \begin{bmatrix}
    \textbf{\textit{B}}_{1} & \textbf{\textit{B}}_{2} \\
    \textbf{\textit{B}}_{2} & \textbf{\textit{0}}
\end{bmatrix}
\end{gather}
where, 
\begin{gather}
\label{eq:11}
 \textbf{\textit{B}}_{1} = -j\frac{1}{R} Re \big\{ \textbf{\textit{Z}}_{AT} \big\}^{-\frac{1}{2}} Im \big\{\textbf{\textit{Z}}_{AT}\big\} Re \big\{\textbf{\textit{Z}}_{AT} \big\}^{-\frac{1}{2}} 
\end{gather} 

\begin{gather}
\label{eq:12}
 \textbf{\textit{B}}_{2} = j\frac{1}{\sqrt{R}} Re\big\{ {\textbf{\textit{Z}}_{AT}}\big\} ^{-\frac{1}{2}}.
\end{gather}

To calculate each element of the DMN-LE presented in Fig.\ref{fig:fig2} the matrix \eqref{eq:2} is used where each of its elements are defined in \eqref{eq:3}. 

The conversion of each element of the matrix $\textbf{\textit{Y}}_{MT}$ into capacitive or inductive elements, is performed based on \eqref{eq:4} and \eqref{eq:5}.

 \begin{gather} \label{eq:2}
 \textbf{\textit{Y}}_{MT}
  =
 \begin{bmatrix}
    y_{11} & y_{12} & y_{13} & y_{14} \\
    y_{12} & y_{22} & y_{23} & y_{24} \\
    y_{13} & y_{23} & y_{33} & y_{34} \\
    y_{14} & y_{24} & y_{34} & y_{44}
\end{bmatrix}
\end{gather}
where, 
\begin{equation}\label{eq:3}
    \begin{split}
    Y_{1} = y_{11} + y_{12} + y_{13} + y_{14} \\ 
    Y_{2} = y_{22} + y_{12} + y_{23} + y_{24} \\
    Y_{3} = y_{33} + y_{13} + y_{23} + y_{34} \\
    Y_{4} = y_{44} + y_{14} + y_{24} + y_{34} \\
    Y_{5} = -y_{12}\\
    Y_{6} = -y_{13}\\
    Y_{7} = -y_{14}\\
    Y_{8} = -y_{23}\\
    Y_{9} = -y_{24}\\
    Y_{10} = -y_{34}
    \end{split}
\end{equation} 
Thus, if $Y_{\textit{i}}$ $\textgreater$ 0, 

\begin{equation}\label{eq:4}
    \begin{split}
    C_i = \frac{Y_{\textit{i}}}{\omega}
    \end{split},
\end{equation} 
and if $Y_{\textit{i}}$ $\textless$ 0, we have, 

\begin{equation}\label{eq:5}
    \begin{split}
    L_i = \frac{1}{Y_{\textit{i}}\omega}
    \end{split},
\end{equation} 
%\frac{1}{R}
where $\omega=2\pi.f_{r}$.

For the DMN-LE modeling with the addition of losses the equivalent circuit model is used in which the concept of the Q-factor is employed. In this way, through the use of the Q-factor obtained from the manufacturer of the electronic component, \cite{Cit14} and \cite{Cit15}, it is possible to design a more realistic structure with losses.

%%%%%%%%%%%%%%%%%%%%%%%%%%%

% ----------------------------------------------------------
% Seção DMN-RH
% ----------------------------------------------------------
\section{DMN with Ring Hybrid (DMN-RH)}

\begin{figure}[h!]
    \centerline{\includegraphics[width=0.5\textwidth]{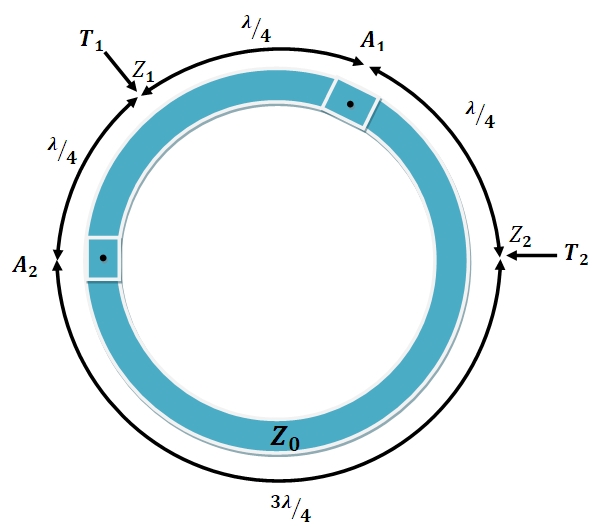}}
    \caption{DMN-RH.}
    \label{fig:fig10}
\end{figure}

The Ring Hybrid (RH) is a linear microwave device which is generally defined as a four-port \cite{Cit11}. It is composed of the branch-line coupler design with a 90$^{\circ}$ hybrid junction, with spacing of $\lambda/4$ between ports $T_1$ and $A_2$, $T_1$ and $A_1$, and $A_1$ and $T_2$. The spacing between ports $T_2$ and $A_2$ is $3\lambda/4$. Such a Ring Hybrid is readily implemented using microstrip lines as shown in Fig.\ref{fig:fig10}. In relation to obtaining the impedance matching of the ports, a transmission line segment is added to each port that one wishes to realize the matching in order to obtain the expected impedance transformation. 

In this way, the RH is a useful tool as a solution for the DMN acting in the decoupling of the elements of an antenna array and impedance matching. Fig.\ref{fig:fig10} shows the standard design of the RH with impedance characteristic $Z_0$. 

\subsection{Design of the DMN-RH}
The RH modeling is based on the formatting of the microstrip configuration parameters that will be used. To the design of DMN-RH it is necessary to define the parameters of the transmission line segments both in the ring itself and at the ports $T_1$ and $T_2$. Thus to define the line widths of the ring the impedance $Z_{0}$ and the port impedances $Z_{1}$ ($T_1$) and $Z_{2}$ ($T_2$) are required. The derivation of these parameters is based on the circuit analysis of the microstrip ring hybrid according to \cite{Cit11}. For the derivation of parameter $Z_0$ (characteristic impedance RH) \eqref{eq:7} is used based on the impedance parameters of the dipole antenna array, the $\textbf{\textit{Z}}_{AT}$ matrix, defined previously. Once $\textbf{\textit{Z}}_{AT}$ is available it is possible to define: $\textit a = Z_{11} = Z_{22}, b = Z_{12} = Z_{21}$ and 
\begin{equation}\label{eq:7}
    \begin{split}
    Z_0 = \sqrt{\sqrt{4 |a+b| |a-b|R^2}}
    \end{split},
\end{equation} 
where $R$ represents the impedance of the sources connected to $T_1$ and $T_2$.

For the derivation of the impedances $Z_{01}$ and $Z_{02}$ it is necessary first to calculate the impedances $Z_1$ and $Z_2$, given by \eqref{eq:6}. Based on these values \eqref{eq:8} derives $Z_{01}$ and $Z_{02}$.

\begin{equation}\label{eq:6}
    \begin{split}
    Z_1 = \frac{{Z_0}^2/ 2}{a+b} \\
    Z_2 = \frac{{Z_0}^2/ 2}{a-b} \\
    \end{split}   ,
\end{equation} 
where,
\begin{equation}\label{eq:8}
    \begin{split}
    Z_{0i} = \sqrt{R.Re\{{Z_i}\} - \frac{R. Im\{Z_i\}^2}{R-Re\{{Z_i}\}}}
    \end{split}.
\end{equation} 

It is important to mention that \eqref{eq:8} is valid only if the value of $Z_{0i}$ obtained is purely real, resulting in a matching impedance with a single transmission line. Otherwise, this matching impedance will not be possible with a single strip line, needing least two or more of them.

For the definition of the physical length of the i-th line segment, the $\theta_{i}$ parameter is given as the electric length $\theta_{i} = \beta l_ {i}$ and derived from \eqref{eq:9}. 

\begin{equation}\label{eq:9}
    \begin{split}
    \tan(\beta l_i) = \frac{(R-Re\{ Z_i\})Z_{0i}}{R.Im\{Z_i \}} \end{split}.
\end{equation} 

% ----------------------------------------------------------
% Seção ND
% ----------------------------------------------------------
\section{Networkless Decoupling and Matching (NDM)}
\begin{figure}[h]
    \centerline{\includegraphics[width=0.3\textwidth]{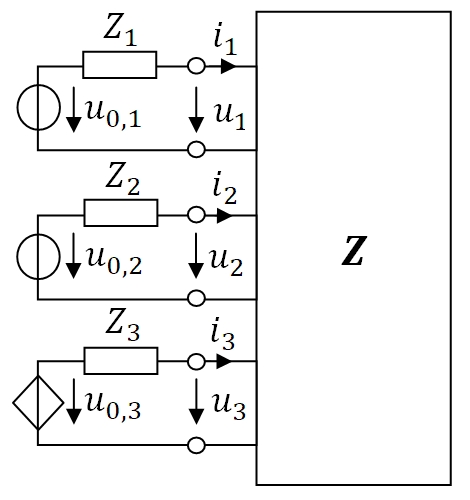}}
    \caption{Three linear generators connected to a linear 3-port.}
    \label{fig:fig17}
\end{figure}

This method aims to achieve the impedance matching and decoupling between the elements of the antenna array, without having a multi-port impedance matching/decoupling structure. For this, a new antenna element is inserted equidistantly between the existing elements. Furthermore, 3 linear voltage sources are employed as shown in Fig.\ref{fig:fig17}.

Each port has its voltage phasor given by \eqref{eq:13}. The generators must also be designed to have proper output impedance $Z_i$.   

\begin{equation}\label{eq:13}
    \begin{split}
    u_i = \frac{u_{0,i}}{1+\frac{Z_i}{Z_i^*}}
     \end{split}
     , i = 1,2,3
\end{equation} 

In summary, the method allows the first two generators to have arbitrary values, while the voltage of the third generator, $u_{0,3}$ is controlled by the other two, according to \eqref{eq:14}, where $g_1$ and $g_2$ are properly chosen complex constants.

\begin{equation}\label{eq:14}
    \begin{split}
    u_{0,3} = g_1u_{0,1}+g_2u_{0,2}
     \end{split}
\end{equation} 

\subsection{Design Networkless Matching}
For the NDM modeling it is necessary to define the vectors $\textbf{\textit{u}}=[u_1 u_2 u_3]^T$, $\textbf{\textit{u}}_0 = [u_{01} u_{02} u_{03}]^{T}$ and $\textbf{\textit{i}}=[i_1 i_2 i_3]^{T}$. $\textbf{\textit{Z}}$ is the impedance matrix of the antenna array consisting of three elements and $\textbf{\textit{Z}}_0$ the diagonal matrix of the three source impedances. 

\begin{gather} \label{eq:17}
  \mathbf{Z} =
  \begin{bmatrix}
        a & b & c \\
        b & a & c \\
        c & c & a 
    \end{bmatrix},
\end{gather}

\begin{gather} \label{eq:16}
  \mathbf{Z}_{0} =
  \begin{bmatrix}
        Z_{1} & 0 & 0 \\
        0 & Z_{2} & 0 \\
        0 & 0 & Z_{3}
    \end{bmatrix}
\end{gather}

with, $a= Z_{11}$, $b= Z_{12}$ and $c= Z_{13}$.

Requiring power matching as given by \eqref{eq:13} and \eqref{eq:14} leads to the following derivation \cite{Cit12}.

\begin{gather} \label{eq:18}
  { (\textbf{1}-{\mathbf{Z}}_0({\mathbf{Z}}_0+{\mathbf{Z}})^{-1})
    \begin{bmatrix}
        1 & 0 \\
       0 & 1 \\
       g_1 & g_2
 \end{bmatrix}
 = 
  \begin{bmatrix}
        x_1 & 0 \\
       0 & x_2 \\
       g_1 x_3 & g_2 x_3
 \end{bmatrix}
  }
\end{gather}

where, 
\begin{equation}\label{eq:19}
    \begin{split}
    x_1 = (1+\exp{(2j\arg Z_1)})^{-1}\\
    x_2 = (1+\exp{(2j\arg Z_2)})^{-1}\\
    x_3 = (1+\exp{(2j\arg Z_3)})^{-1}.
    \end{split}
\end{equation} 

Thus, given $\textbf{\textit{Z}}$ we solve the above equation for $\textbf{\textit{Z}}_0$, $g_1$ and $g_2$ to have,
\begin{equation}\label{eq:20}
    \begin{split}
    g_{1} = g_{2} = 
    \frac{{-c}^{2}+(a+Z_3)b}{(b-a+Z_1)c}
    \end{split}
\end{equation} 

\begin{equation}\label{eq:21}
    \begin{split}
    Z_{1} = Z_{2} = a^{*}-b^{*}
     \end{split}
\end{equation} 

\begin{equation}\label{eq:22}
    \begin{split}
    Z_{3} = a^{*}-(\frac{{c^2}}{b})^*
     \end{split}.
\end{equation} 

Finally we have to transform the impedances $Z_1$, $Z_2$ and $Z_3$ to the source resistance of $50\Omega$ by impedance matching twoports (see Fig.\ref{fig:fig18}).

For analyzing the NDM circuit voltage controlled voltage sources (VCVS) are used. The values of $\textbf{\textit{u}}_0$ are the voltages of the sources as defined in Fig.\ref{fig:fig17}, while the voltages $\textbf{\textit{u}}_0'$ are scaled to compensate for the effect of the aforementioned impedance matching twoports. In this way the value used as gain ($G$) of the VCVS is set by the values of $\textbf{\textit{u}}_0'$.

% ---
% Resultados
% ---

\section{Numerical Results}
Numerical results were obtained to enable a comparison of the performance of the three concepts in terms of the S parameters. The following simulation conditions are assumed:

\begin{itemize}
\item Two-element antenna array based on half-wavelength dipole \( L = \lambda/2 \).
\item Distance between antennas: \( d = \lambda/4 \).
\item Reference carrier frequency: $f_r =  3 GHz$.
\item Impedance Matrix with a classical model is computed according to \cite{Cit13}.
\item Frequency Range of interest (for measuring bandwidth): 2.9 GHz - 
3.1GHz.
\end{itemize}
The main objective of this analysis is to verify the behavior of the mutual coupling ($S_{12}$ and $S_{21}$) and impedance matching ($S_{11}$ and $S_{22}$) of the proposed DMN approaches over the frequency range of interest. Circuit simulations employ the ANSOFT DESIGNER SV software. 

%%%%%%%%%
%%%%%%%%%%%%%%%%%%%%%%%%
%%%%%%%%
\subsection{Baseline Performance}
\begin{figure}[h!]
    \centering
    \includegraphics[width=\columnwidth]{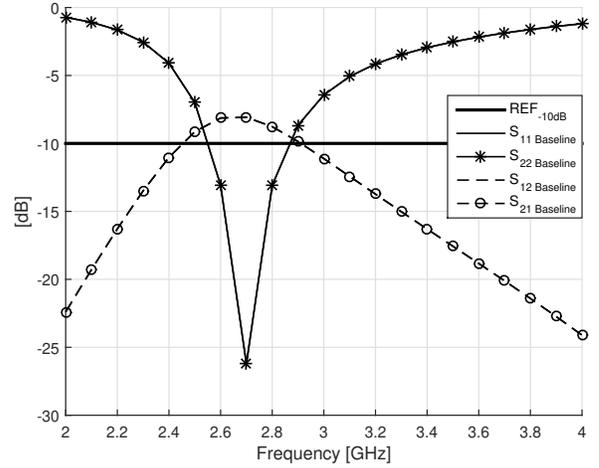}
    \caption{Baseline Array model.}
    \label{fig:fig4}
\end{figure}

The performance of the array antenna without DMN is presented in Fig.\ref{fig:fig4}. It shows that the resonance frequency performance is around 2.7 GHz (i.e 300 MHz off the reference frequency). Impedance matching ($S_{11}$ or $S_{22}$), has value of around -26 dB, at resonance frequency, while for the reference frequency a value of -6 dB is observed. $S_{12}$ and $S_{21}$ assume a value of around -12 dB at the reference frequency. A reference level of -10 dB is considered for measuring antenna bandwidth from S parameters.

%%%%%%%%%%%%%%%%%%%%%%%%%%%      %%%%%%%%%%%%%%%%%%%%%%%%%%%
%%%%%%%%%%%%%%%%%%%%%%%%%%%      %%%%%%%%%%%%%%%%%%%%%%%%%%%
%%%%%%%%%%%%%%%%%%%%%%%%%%%      %%%%%%%%%%%%%%%%%%%%%%%%%%%

\subsection{Results for DMN-LE}
\begin{figure}[h!]
    %\centering
    \centerline{\includegraphics[width=\columnwidth]{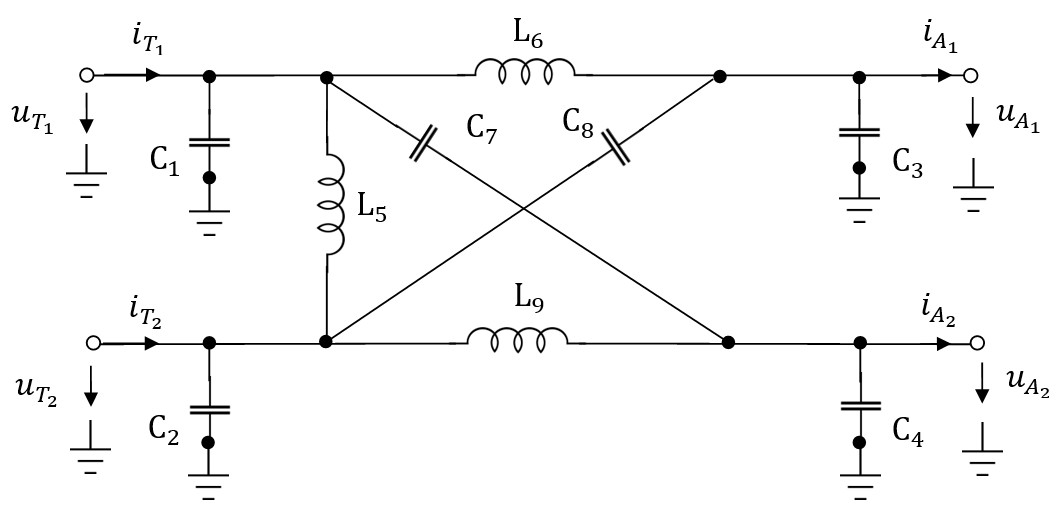}}
    \caption{Circuit DMN-LE.}
    \label{fig:fig16}
\end{figure}

The circuit equivalent of the DMN using lumped elements is shown in Fig.\ref{fig:fig16}. The values of the components are described in Table \ref{tab:table0} in which the values of Q-factor are given according to \cite{Cit14} and \cite{Cit15}.

\begin{figure}[h!]
     %\centering
     \begin{subfigure}[b]{0.485\textwidth}{}
         \includegraphics[width=\columnwidth]{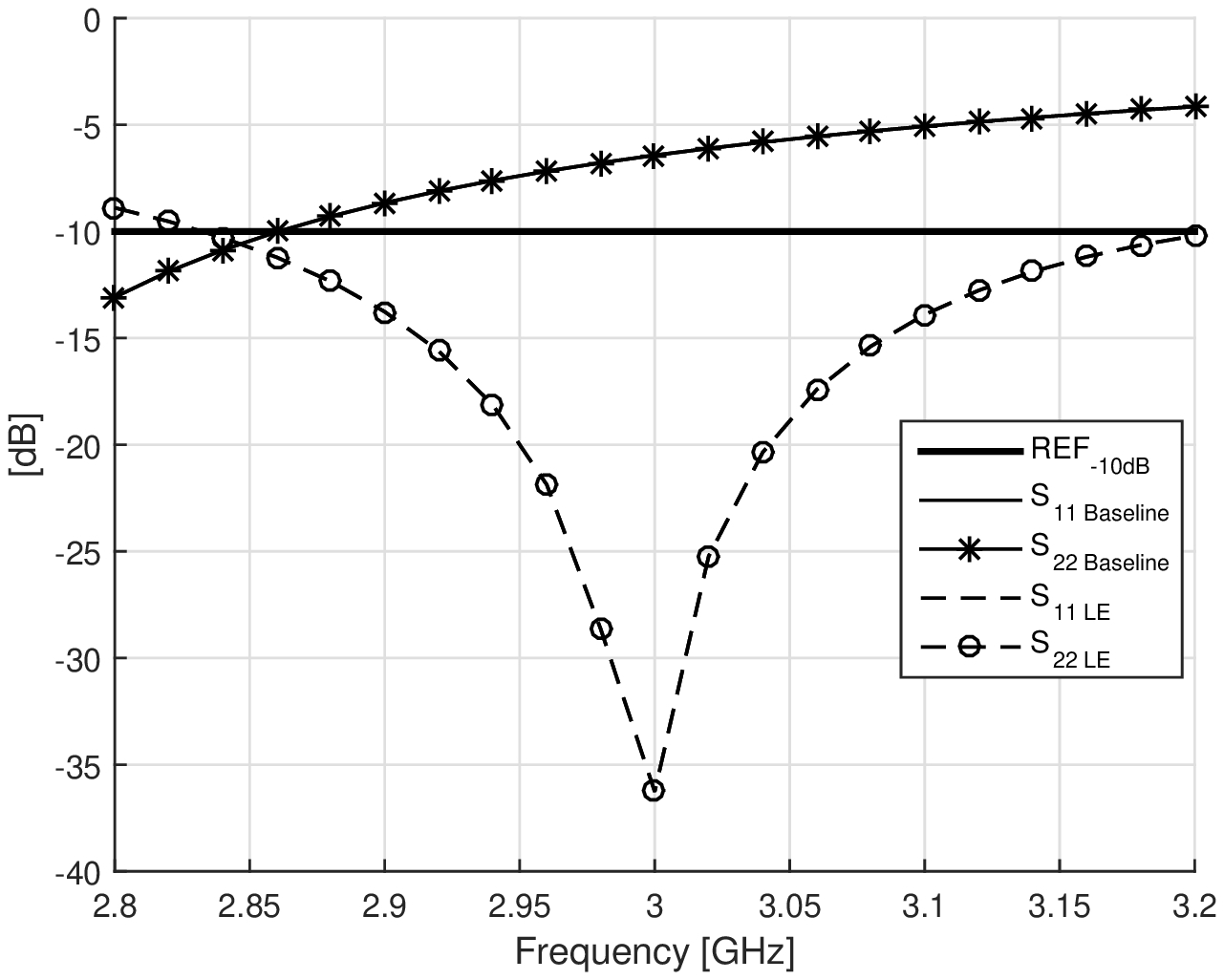}
         \caption{}
         \label{fig:fig5}
     \end{subfigure}
     
    %\begin{subfigure}[b]{0.25\linewidth}
    
     \hfill
     \begin{subfigure}[b]{0.485\textwidth}
         \includegraphics[width=\columnwidth]{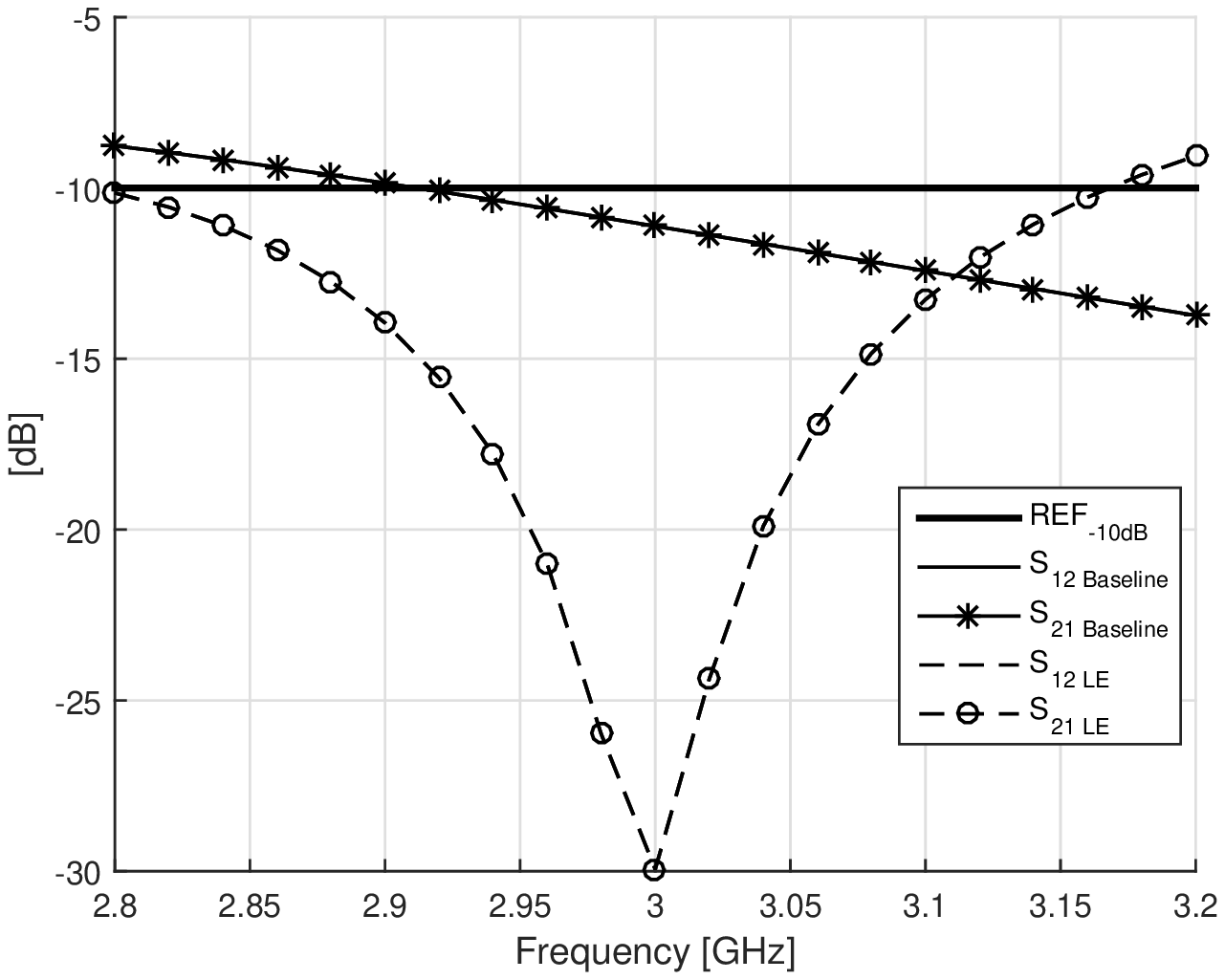}
         \caption{}
         \label{fig:fig6}
     \end{subfigure}
     \hfill
        \caption{DMN-LE versus Baseline Array.}
        \label{fig:2 graphs}
\end{figure}

\begin{table}[h!]
  \begin{center}
    \caption{Values of Components}
    \label{tab:table0}
    \begin{tabular}{c c}
    \textbf{Components} & \textbf{Values Obtained}\\
    \hline
    \hline
    $C_1$, $C_2$ & {0.53888pF (Q = 137, $f_Q$ = 2.4GHz)} \\
    \hline
    $C_3$, $C_4$ & {0.63942pF, (Q = 131, $f_Q$ = 2.4GHz)} \\
    \hline
    $L_5$ & {2.3544nH, (Q = 64.4, $f_Q$ = 2.4GHz)} \\
    \hline
    $L_6$, $L_9$  & {2.7827nH, (Q = 78.9, $f_Q$ = 2.4GHz)} \\
    \hline
    $C_7$, $C_8$ & {0.29575pF, (Q = 146, $f_Q$ = 2.4GHz)} \\
    \end{tabular}
  \end{center}
\end{table}
The behavior of the DMN-LE versus baseline is presented in the Fig.\ref{fig:fig5}, for $S_{11}$ and $S_{22}$. Fig.\ref{fig:fig6} shows the behavior of $S_{12}$ and $S_{21}$. 

For having both the decoupling and the matching better than -20 dB a bandwidth of 100 MHz and for better than -10 dB a bandwidth of more than 300 MHz has been achieved. 

%%%%%%%%%%%%%%%%%%%%%%%%

\subsection{Results for DMN-RH}

The values obtained from the equations presented for DMN-RH are shown in Table \ref{tab:table1}. Based on these values the DMN-RH modeling is performed and the microstrip parameters are obtained. 

\begin{table}[h!]
  \begin{center}
    \caption{DMN-RH Calculated Parameters.}
    \label{tab:table1}
    \begin{tabular}{l|c} % <-- Alignments: 1st column left, 2nd middle and 3rd right, with vertical lines in between
    \textbf{Parameters} & \textbf{Values Obtained}\\
    \hline
        $Z_0$ & {97.1845$\Omega$} \\
        $Z_1$ & {(40.8666 -j5.0754)$\Omega$}\\
        $Z_2$  & {(25.2097 -j55.2266)$\Omega$ }   \\
        $Z_{01}$ & {43.6155$\Omega$}\\
        $\theta_{01}$ & {-57.5014$^{\circ}$ (122.498$^{\circ}$)}\\
        $Z_{02}$ & {j69.936$\Omega$}\\
    \end{tabular}
  \end{center}
\end{table}

To obtain the parameters of width and length from the impedance value and the electric length, Ansoft Design SV's microstrip line calculator, [16], is used. To this modelling, the substrate of the microstrip line RO3006 material is used \cite{Cit17} which has $\epsilon_r = 6.15$ and thickness of the substrate($h$) is $1.52$mm. The thickness of the strip line and the ground plane($t$) is $35 \mu$m.

In this design it is important to note that $Z_{02}$ has purely imaginary impedance value. In this way, it is not possible to calculate the length and width values of the microstrip line of the port from these equations.

Two possible solutions are given for matching port $T_2$ to $50 \Omega$. In the first solution the reactive part of $Z_2$ is eliminated by a strip line with $Z_{21} = 50 \Omega$ and appropriate length. Then a strip line with length $\lambda/4$ and appropriate $Z_{22}$ is used to provide matching to a $50 \Omega$ source. This procedure results in the values reported in Table \ref{tab:table2}.

\begin{table}[h!]
  \begin{center}
    \caption{DMN-RH Design Parameters.}
    \label{tab:table2}
    \begin{tabular}{l|c} % <-- Alignments: 1st column left, 2nd middle and 3rd right, with vertical lines in between
    \textbf{Parameters} & \textbf{Values Obtained}\\
    \hline
    \hline
        $\theta_{r}$ & {$90^{\circ}$}\\
        $l_r$ & {12.5859 mm} \\
        $w_r$ & {0.4356 mm} \\
        \hline
        $\theta_{1}$ & {$122.498^{\circ}$}\\
        $l_{1}$ & {15.8662 mm}\\
        $w_{1}$ & {2.7999 mm}\\
        \hline
        $Z_{21}$ & {$50\Omega$}\\
        $\theta_{21}$ & {$51.056^{\circ}$}\\
        $l_{21}$ & {6.7001 mm}\\
        $w_{21}$ & {2.2016 mm}\\
        \hline
        $Z_{22}$ & {$23.3544\Omega$}\\
        $\theta_{22}$ & {$90^{\circ}$}\\
        $l_{22}$ & {11.0431 mm}\\
        $w_{22}$ & {7.0753 mm}\\
    \end{tabular}
  \end{center}
\end{table}

\begin{figure}[h!]
    \centerline{\includegraphics[width=\columnwidth]{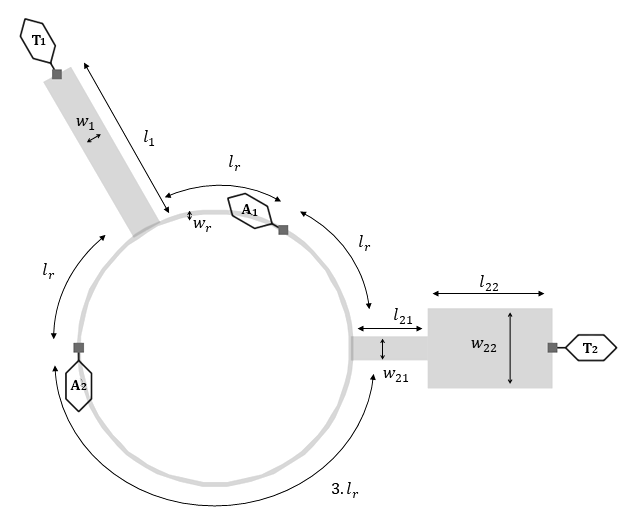}}
    \caption{DMN-RH Design.}
    \label{fig:fig11}
\end{figure}

The complete design of the DMN-RH is shown in the Fig.\ref{fig:fig11}, in which the width ($w_{r}$) of the ring and its length($l_{r}$), the width ($w_{1}$) of the port $T_1$ and its length ($l_{1}$) and for port $T_2$, the widths $w_{21}$, $w_{22}$, and its lengths $l_{21}$, $l_{22}$ are indicated. 

It should be noted that for the line that above length $l_{21}$ and its length is configured as a transmission line with a characteristic impedance $R$.
It should be noted that although the electrical lengths ($\theta_r$ and $\theta_{22}$) of the microstrip lines are equal the impedance values of each line are different consequently producing both the parameters of different physical length and width.

The second alternative as a solution for impedance matching of the port $T_2$ is the use of stubs to eliminate the imaginary part of $Z_2$ and after this procedure a microstrip line of 90$^{\circ}$ is used to realize the impedance matching for $R$. 

The impedance of the short circuited stub with an electric length of $\theta = \pi/4$ is given by \eqref{eq:29}. 

\begin{equation}\label{eq:29}
    \begin{split}
    Z_{s_1} = \frac{1}{Im\{Z_{2}\}} 
     \end{split}
\end{equation} 

\begin{figure}[h!]
    \centerline{\includegraphics[width=\columnwidth]{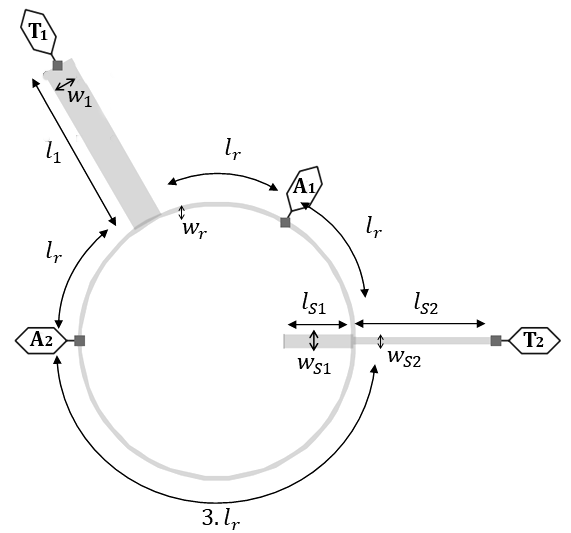}}
    \caption{DMN-RH-Stub Design.}
    \label{fig:fig122}
\end{figure}

The Tab.\ref{tab:table21} shows the parameters for modeling of the stub structure and strip line of the port $T_2$. The results of this modeling will be called DMN-RH-Stub and is shown in the Fig.\ref{fig:fig122}.

\begin{table}[h!]
  \begin{center}
    \caption{DMN-RH-Stub Design Parameters.}
    \label{tab:table21}
    \begin{tabular}{l|c}
    \textbf{Parameters} & \textbf{Values Obtained}\\
    \hline
    \hline
        $Z_{S1}$ & {$66.7342\Omega$}\\
        $\theta_{S1}$ & {45$^{\circ}$}\\
        $l_{S1}$ & {6.0741 mm}\\
        $w_{S1}$ & {1.2249 mm}\\
        \hline
        $Z_{S2}$ & {$85.44\Omega$}\\
        $\theta_{22}$ & {90$^{\circ}$}\\
        $l_{S2}$ & {12.4345 mm}\\
        $w_{S2}$ & {0.6503 mm}\\
    \end{tabular}
  \end{center}
\end{table}

\begin{figure}[h]
     \centering
     \begin{subfigure}[b]{0.485 \textwidth}
         \centering
         \includegraphics[width=\columnwidth]{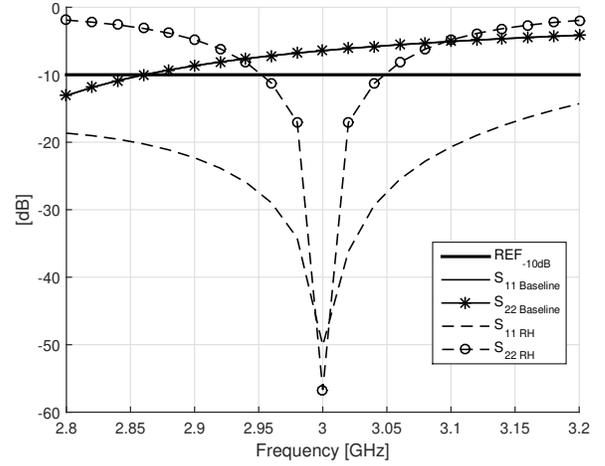}
         \caption{}
         \label{fig:fig12}
     \end{subfigure}
     \hfill
     \begin{subfigure}[b]{0.485 \textwidth}
         \centering
         \includegraphics[width=\columnwidth]{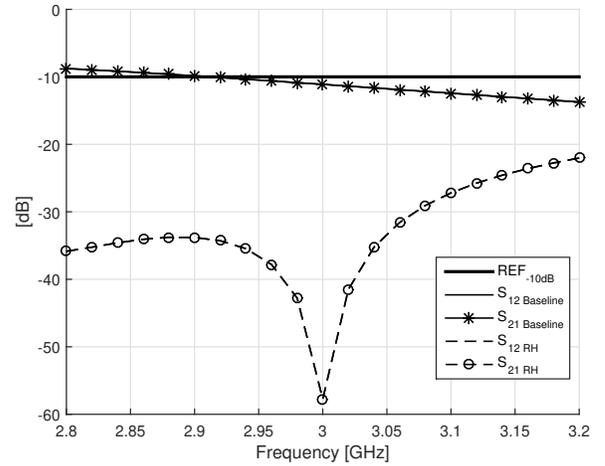}
         \caption{}
         \label{fig:fig13}
     \end{subfigure}
     \hfill
        \caption{DMN-RH versus Baseline Array}
        \label{fig:12 graphs}
\end{figure}

\begin{figure}[h]
     \centering
     \begin{subfigure}[b]{0.485 \textwidth}
         \centering
         \includegraphics[width=\columnwidth]{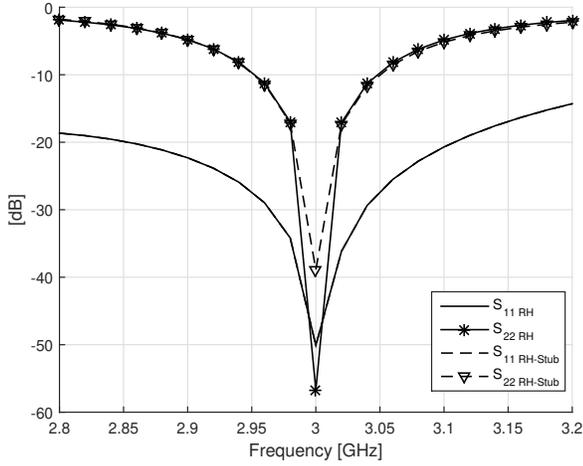}
         \caption{}
         \label{fig:fig12st}
     \end{subfigure}
     \hfill
     \begin{subfigure}[b]{0.485 \textwidth}
         \centering
         \includegraphics[width=\columnwidth]{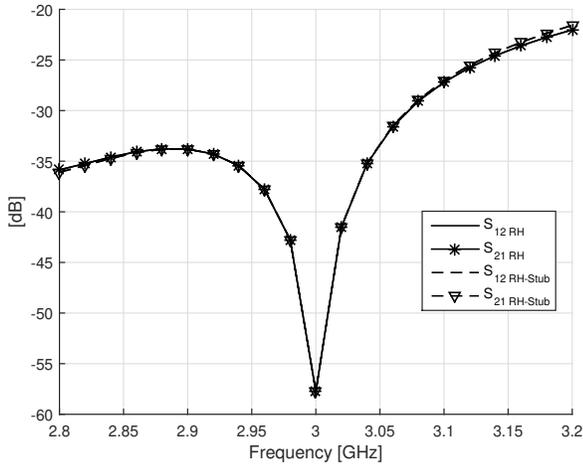}
         \caption{}
         \label{fig:fig13st}
     \end{subfigure}
     \hfill
        \caption{DMN-RH versus DMN-RH-Stub}
        \label{fig:13 Stub graphs}
\end{figure}

As a result of the analysis of the DMN-RH, the parameters $S_{11}$, $S_{12}$, $S_{21}$ and $S_{22}$ are presented in the Figs. \ref{fig:fig12}, \ref{fig:fig13}, \ref{fig:fig12st} and \ref{fig:fig13st}.

The performance of the two alternative solutions is quite similar. The bandwidth for decoupling ($S_{21}$ and $S_{12}$) better than -20dB is larger than 400 MHz, for matching at port $T_1$ ($S_{11}$) better than -20dB is 250 MHz, while matching at port $T_2$ ($S_{22}$) is the bottleneck. Here a bandwidth of only 100 MHz for being better than -10dB and less than 50 MHz for better than -20dB is achieved.

%%%%%%%%%%%%%%%%
%%%%%%%%%%%%%%%% 
%%%%%%%%%%%%%%
%%%%%%%%%%%%%%%
\subsection{Results for Networkless Decoupling and Matching}
\begin{figure}[h!]
    \centering
    \centerline{\includegraphics[width=0.5\textwidth]{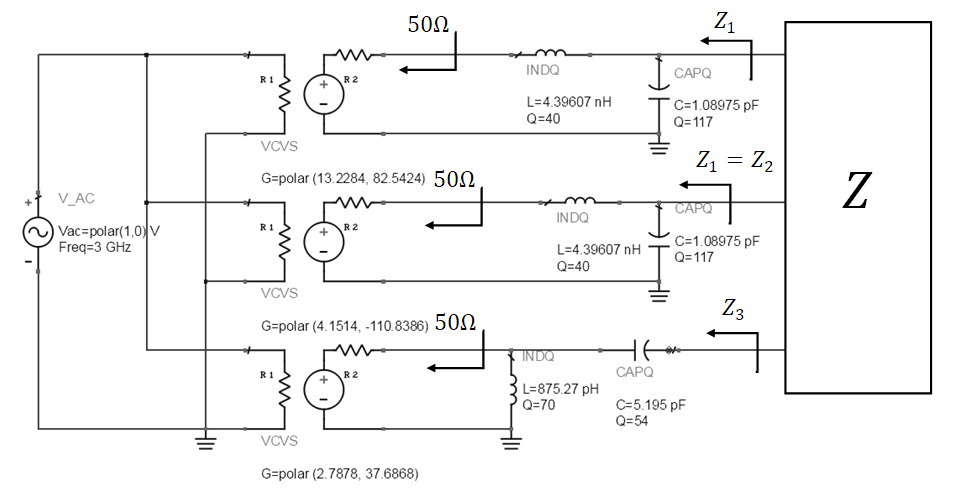}}
    \caption{NDM Circuit.}
    \label{fig:fig18}
\end{figure}

The results obtained from \eqref{eq:13} to \eqref{eq:22} for the NDM modeling are presented in the Tables \ref{tab:table2829} and \ref{tab:table3}, as well as can be visualized in the Fig.\ref{fig:fig18}, in which NDM analysis circuit is presented. It is formed by an AC voltage source, defined with a polar voltage of modulo $1V$ and phase $0^{\circ}$, and connected to three VCVS each instantiated by the respective values of $\textbf{\textit{u}}_{0}^{'}$. 

The VCVSs are connected to LC-twoports, which provide the impedance matching from $50 \Omega$ to the required impedances $Z_{1}$, $Z_{2}$ and $Z_{3}$ given in Table \ref{tab:table2829}. The conversion from $\textbf{\textit{u}}_{0}$ to $\textbf{\textit{u}}_{0}^{'}$ compensates for the magnitude and phase change caused by these matching twoports.   

\begin{table}[h]
  \begin{center}
    \caption{Values for $\textbf{\textit{Z}}$ and $\textbf{\textit{Z}}_o$}
    \label{tab:table2829}
    \begin{tabular}{l|c}
    \textbf{Parameters} & \textbf{Values Obtained ($\Omega$)}\\
    \hline
    \hline
        $a$ & {$73.05+j42.44$}\\
        $b$ & {$40.74-j28.31$} \\
        $c$ & {$64.11-j0.074$} \\
        \hline
        $Z_{1}$ & {$32.3-j70.76$}\\
        $Z_{2}$ & {$32.3-j70.76$}\\
        $Z_{3}$ & {$4.09+j4.66$}\\
    \end{tabular}
  \end{center}
\end{table}

\begin{table}[h]
  \begin{center}
    \caption{NDM Design Parameters.}
    \label{tab:table3}
    \begin{tabular}{l|c} % <-- Alignments: 1st column left, 2nd middle and 3rd right, with vertical lines in between
    \textbf{Parameters} & \textbf{Values Obtained} (V)\\
    \hline
    \hline
        $u_{0,1}$ & {$10.6327\mathrm{e}^{-j41.8153}$}\\
        $u_{0,2}$ & {$3.368\mathrm{e}^{ j124.8037}$} \\
        $u_{0,3}$ & {$0.8736\mathrm{e}^{ j109.4254}$} \\
        \hline
        $u_{0,1}^{'}$ & {$13.2283\mathrm{e}^{ j82.5424}$}\\
        $u_{0,2}^{'}$ & {$4.1514\mathrm{e}^{-j110.8386}$}\\
        $u_{0,3}^{'}$ & {$2.7878\mathrm{e}^{ j37.6868}$}\\
        \hline
        $g_{1}, g_{2}$ & {$0.1176\mathrm{e}^{j145.2833}$}\\
    \end{tabular}
  \end{center}
\end{table}

\begin{figure}[h!]
     \centering
     \begin{subfigure}[b]{0.485 \textwidth}
         \centering
         \includegraphics[width=\columnwidth]{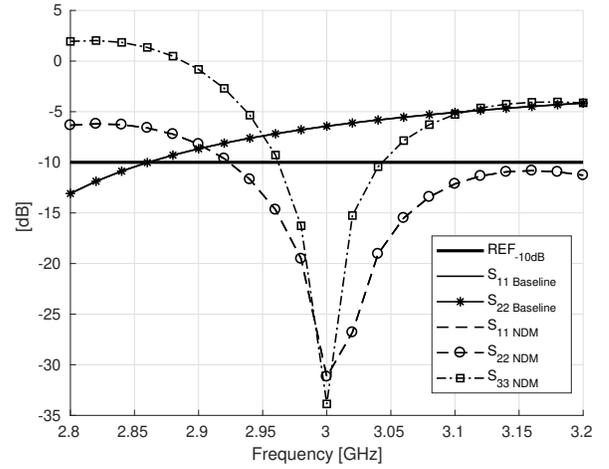}
         \caption{}
         \label{fig:fig14}
     \end{subfigure}
     \hfill
     \begin{subfigure}[b]{0.485 \textwidth}
         \centering
         \includegraphics[width=\columnwidth]{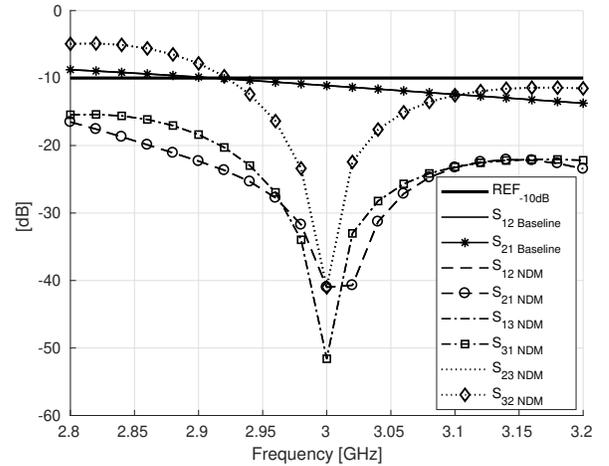}
         \caption{}
         \label{fig:fig15}
     \end{subfigure}
     \hfill
        \caption{NDM versus Baseline Array}
        \label{fig:15 graphs}
\end{figure}

Fig.\ref{fig:15 graphs} shows that the bandwidth for decoupling being better than -20dB is quite large (more than 250 MHz), for matching the achievable bandwidth is less than 100 MHz for better than -20dB and more than 200 MHz for -10dB. 

\subsection{Comparison Among DMN approaches}
\begin{figure}[h!]
     \centering
     \begin{subfigure}[b]{0.485 \textwidth}
         \centering
         \includegraphics[width=\textwidth]{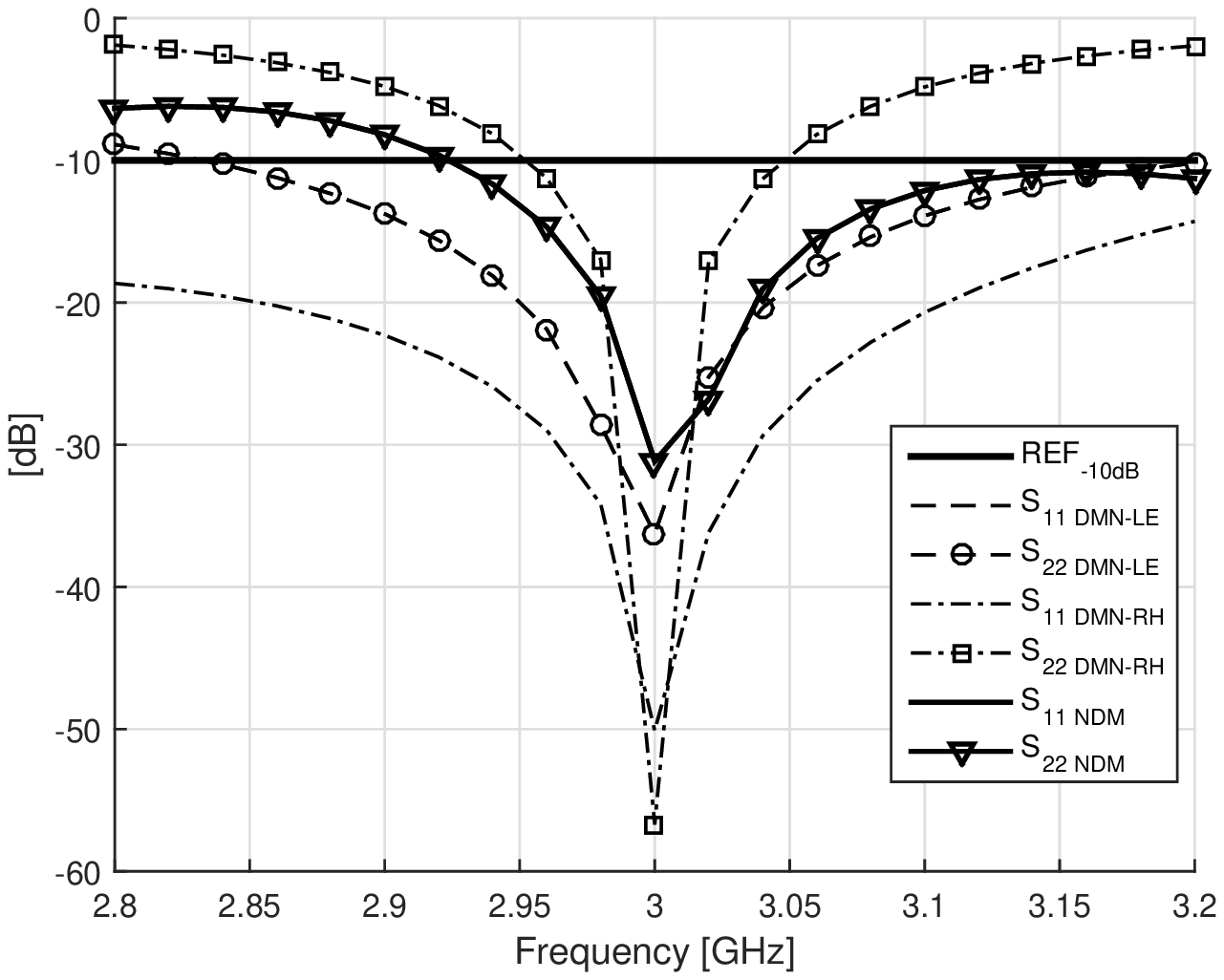}
         \caption{}
         \label{fig:fig19}
     \end{subfigure}
     \hfill
     \begin{subfigure}[b]{0.485 \textwidth}
         \centering
         \includegraphics[width=\textwidth]{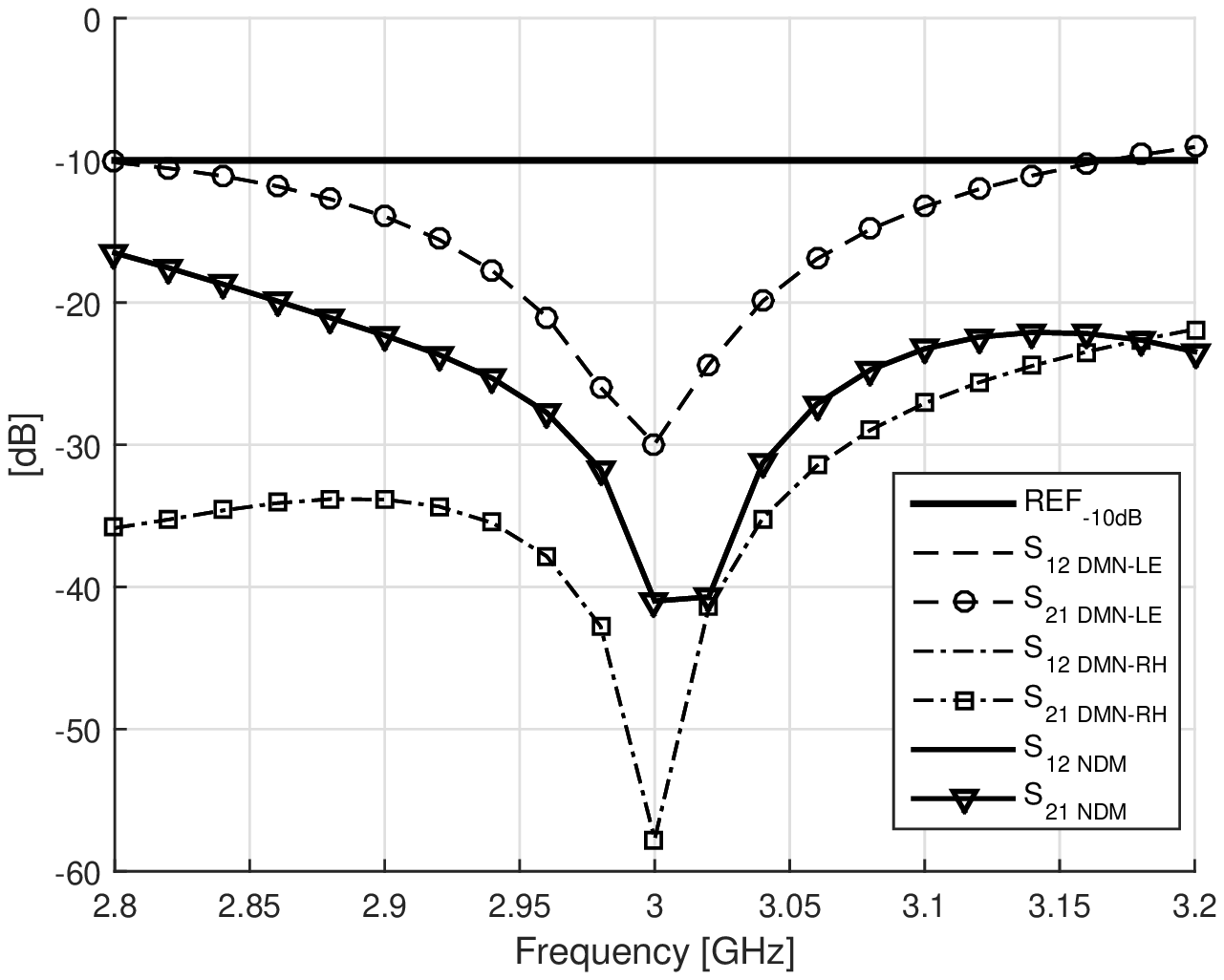}
         \caption{}
         \label{fig:fig20}
     \end{subfigure}
     \hfill
        \caption{Comparison Analysis Parameters - $S_{11}$, $S_{22}$, $S_{12}$ and $S_{21}$}
        \label{fig:20 graphs}
\end{figure}

Fig.\ref{fig:20 graphs} shows the graph of comparison between each one of the three DMN proposed methods. In this way it is possible to perceive the behavior improvement in each method. The NDM method in turn stands out for having the best improvement in relation to the reference frequency for $S_{11}$ and $S_{12}$, since it is analyzed the behavior in both ports and also being able to be considering issues as the compact structure and modeling in relation to the others.

Analyzing the results it is possible to infer that there were significant improvements with the inclusion of DMN techniques compared to the baseline array. In this way, the NDM presents an intermediate performance in terms of $S_{11}$ and $S_{22}$ in relation the two other presented models, looking for a middle ground between the model with Lumped Elements and the Ring Hybrid making it be defined as an excellent alternative of application.
Despite this, the bandwidth of this method, considering the reference threshold of -10dB, is around at the limit of the frequency range interest, providing a value around 200 MHz bandwidth. 

As for DMN-RH, it can be concluded that the different behaviors between ports $T_1$ and $T_2$ reduce its bandwidth, forcing the designer to resort to other bandwidth expansion techniques. The DMN-LE, in turn, has a good bandwidth behavior, but its main disadvantage is the circuit complexity might became too cumbersome.

For the behavior of $S_{12}$ and $S_{21}$ NDM also showed a excellent performance in the reference frequency as well as a value below -20dB in the range from 2.9 GHz to 3.1 GHz, featuring good port decoupling.

% ---
% Conclusão
% ---
\section{Conclusion}
The analysis carried out in this paper describes three methods for impedance matching and decoupling for a baseline antenna array in which the scattering parameters ($S$) are compared. For this, a theoretical reference of each method was presented describing its modeling and detailing its parameters.

In this way, the behavior of $S_{11}$, $S_{22}$, $S_{12}$ and $S_{21}$ was investigated for the 2.9 GHz to 3.1 GHz frequency band (200 MHz bandwidth) around the reference frequency of 3 GHz.

With the DMN-LE method, it can be observed that despite the favorable behavior with respect to the baseline, it demands a complex circuit structure which makes its design difficult for compact applications.

The DMN-RH method presented a different performance in terms of bandwidth for each of its ports which reduces its overall bandwidth. Therefore, it ends up being discarded in favor of a DMN solution that offers the same behavior in both output ports.

The NDM method was shown to perform excellent, enabling a compact and efficient antenna structure for both the impedance matching and the minimization of the destructive effects of coupling between the elements of the proposed antenna array.

Therefore, for the overall conclusion, it is possible to notice that from a impedance matching point of view, the DMN-LE provides the highest bandwidth, while, from a decoupling point of view, the DMN-RH is the best and the NDM is between the two aspects. So the choice for each method will depend on the specific requirements, in which one of the three best suits.

\end{document}